\documentclass[twocolumn,preprintnumbers,amsmath,amssymb]{revtex4}
\usepackage{graphicx} % include figure files
\usepackage{setspace} % include figure files
\usepackage{dcolumn} % align table columns on decimal point
\usepackage{bm} % bold math
\usepackage{natbib}
\usepackage{hyperref}
\hypersetup{colorlinks=true,linkcolor=blue,citecolor=blue,urlcolor=blue}
\usepackage{epsfig}
\usepackage{amsmath}
\begin{document}
\title{Effect of crystal field engineering and Fermi level optimization on thermoelectric properties of Ge$_{1.01}$Te: Experimental investigation and theoretical insight}
\author{Ashutosh Kumar$^{1,@,}$\footnote{Email: ashutosh.kumar@universite-paris-saclay.fr}}
\author{Preeti Bhumla$^{2,@}$}
\author{D. Sivaprahasam$^3$}
\author{Saswata Bhattacharya$^{2,}$\footnote{Email: saswata@physics.iitd.ac.in}}
\author{Nita Dragoe$^{1,}$\footnote{Email: nita.dragoe@universite-paris-saclay.fr\\
\\
@ Equal contribution}}

\affiliation{$^1$ICMMO (UMR CNRS 8182), Université Paris-Saclay, F-91405 Orsay, France}
\affiliation{$^2$Department of Physics, Indian Institute of Technology Delhi, New Delhi 110016, India}
\affiliation{$^3$CAEM, ARCI, IIT Madras Research Park, Taramani, Chennai – 600 113, India
}
\date{\today}
\begin{abstract}
This study shows a method of enhancing the thermoelectric properties of GeTe-based materials by Ti and Bi co-doping on cation sites along with self-doping with Ge via simultaneous optimization of electronic (via crystal field engineering, and precise Fermi level optimization) and thermal (via point-defect scattering) transport properties. The pristine GeTe possesses high carrier concentration ($n$) due to intrinsic Ge vacancies, low Seebeck coefficient ($\alpha$), and high thermal conductivity ($\kappa$). The Ge vacancy optimization and crystal field engineering results in an enhanced $\alpha$ via excess Ge and Ti doping, which is further improved by band structure engineering through Bi doping. As a result of improved $\alpha$ and optimized Fermi level (carrier concentration), an enhanced power factor ($\alpha^2\sigma$) is obtained for Ti--Bi co-doped Ge$_{1.01}$Te. These experimental results are also evidenced by theoretical calculations of band structure, and thermoelectric parameters using density functional theory and Boltztrap calculations, respectively. A significant reduction in the phonon thermal conductivity ($\kappa_{ph}$) from $\sim$ 3.5 W·m$^{-1}$·K$^{-1}$ to $\sim$ 1.06 W·m$^{-1}$·K$^{-1}$ at 300\,K for Ti--Bi co-doping in GeTe, attributed to point-defect scattering due to mass and strain field fluctuation, in line with the Debye-Callaway model. The phonon dispersion calculations show a decreasing group velocity in Ti--Bi co-doped GeTe, supporting the obtained reduced $\kappa_{ph}$. The strategies used in the present study can significantly increase the effective mass, optimize the carrier concentration, and decrease phonon thermal conductivity while achieving an impressive maximum zT value of 1.75 at 773\,K and average zT (zT$_{av}$) of 1.03 for Ge$_{0.91}$Ti$_{0.02}$Bi$_{0.08}$Te over a temperature range of 300-773\,K.\\
\end{abstract}
\maketitle
\section{Introduction}
Thermoelectric (TE) materials exhibit distinctive importance for power generation and solid-state cooling owing to their capability of reversible conversion of heat into electricity without any moving parts and are emerging solutions for the thermal management and energy efficiency caused by the ubiquity of waste heat in modern technological time.\cite{R1, R2} The efficacy of a TE device is assessed by a dimensionless quantity, the figure of merit (zT), which has a dependence on the following physical parameters: Seebeck coefficient ($\alpha$), electrical conductivity ($\sigma$) and total thermal conductivity ($\kappa$=$\kappa_{e}$+$\kappa_{ph}$) having electronic ($\kappa_e$) and lattice ($\kappa_{ph}$) contribution to it and Temperature (T), as zT=$\alpha^2\sigma$T/($\kappa$=$\kappa_{e}$+$\kappa_{ph}$). It is seen that achieving a high zT in a TE material is daunting due to the intertwined relations between these TE parameters. Several innovative strategies have been deployed to decouple electron and phonon transport in material to achieve an elevated zT that includes band-structure engineering,\cite{R4, R5, R6, R7, R8} nanostructuring,\cite{R8a} composite approach,\cite{R8b, R8c, R8d, R8dd} high-entropy concept,\cite{R8e, R8f} etc. In other words, concurrent improvement in electronic transport and prohibition of phonon propagation is the utmost criteria to have high-performance TE materials like PbTe,\cite{R9} SnSe,\cite{R10} skutterudites,\cite{R10a} half-Heusler compounds,\cite{R10b} etc.\\
Doped semiconductors are promising for TE properties due to their suitable electronic structure. GeTe-based materials are one of them showing their applicability in mid-temperature range applications; however, large p-type carrier concentration ($\sim$10$^{21}$ cm$^{-3}$) at 300\,K in GeTe stemming from intrinsic Ge results in small $\alpha$ and high $\kappa_e$ which eventually results in inferior TE performance.\cite{R11} Also, large energy separation ($\Delta$E) between light and heavy valence bands limits to elevate $\alpha$. Further, room temperature $\kappa_{ph}$ for pristine GeTe ($\sim$ 3 W·m$^{-1}$·K$^{-1}$) is significantly higher than the theoretical minimum value (0.44 W·m$^{-1}$·K$^{-1}$) estimated using Cahill's model.\cite{R12}\\
Since the absolute value of $\alpha$ decreases whereas $\sigma$, and $\kappa_e$ increases with carrier concentration ($n$), therefore optimizing $n$ is one of the important steps followed by engineering electronic and phonon band structures to achieve desired TE parameters. A reduction in $\kappa_{ph}$ can be realized via hierarchical architecture engineering,\cite{R13} defect engineering,\cite{R14} composite approach,\cite{R15} and other multi-scale scattering centers approach.\cite{R17, R18} Several innovative strategies including alloying, and band structure modifications, have been adopted to reduce the high hole carrier concentration and improve $\alpha$ in GeTe.\cite{R19, R20, R21} Bi$^{3+}$, Sb$^{3+}$ aliovalent doping at the Ge$^{2+}$ site have been employed to reduces the carrier concentration and decrease $\kappa_{ph}$ due to phonon scattering through solid-solution point defects.\cite{R22} Also, such aliovalent doping converges the $\Delta$E between light and heavy bands and hence improves $\alpha$. However, large aliovalent doping reduces the $\sigma$ significantly. On the other hand, transition metal doping (such as Ti, Zn, Mn, etc) has been adopted in literature to converge the $\Delta$E to achieve enhanced $\alpha$.\cite{R23, R24, R25}\\
As mentioned earlier, the optimization of the Fermi level (carrier concentration) is of utmost importance to achieve higher zT in a TE material.\cite{R26} For pristine GeTe, extremely high $n$ owing to intrinsic Ge vacancies pushes the Fermi level deep in the valence band. Shuai et. al., adopted the Ge vacancy manipulation to achieve high zT in Ge-rich GeTe system.\cite{R27} Herein, we deploy multiple strategies to achieve a significant improvement of zT (1.75 at 773\,K) via systematic doping of Ti and Bi in vacancy engineered Ge$_{1.01}$Te assisted by the confluence effect of crystal field engineering, valence band convergence, and point defect scattering. Excess Ge manipulates the $n$ which enhances $\alpha$ and reduces $\kappa_e$ at 300\,K. Ti doping further improves $\alpha$ due to crystal field engineering due to a decreased $c/a$ ratio. The Ti--Bi co-doping further improves the band convergence and elevates $\alpha$ and reduces $\kappa_{ph}$ owing to point defect scattering. Further, a theoretical understanding of crystal field and band structure engineering for electronic transport followed by the calculation of TE parameters using the Boltzmann transport equation has been provided. Phonon dispersion calculations further support the phonon engineering-mediated reduced lattice thermal conductivity in the present study.
\section{Methods}
\subsection{Experimental Details}
Ge$_{1.01-x-y}$Ti$_{x}$Bi$_{y}$Te (0$\leq$\textit{x}$\leq$0.02, 0$\leq$\textit{y}$\leq$0.08) was synthesized\textit{via} direct melting of high purity Ge, Ti, Bi, and Te ($>$  99.99\%, Alfa Aesar) in the stoichiometric amount in evacuated quartz ampoules (10$^{-6}$ mbar). The sealed ampoules were heated at 1173\,K for 8 hours with a heating rate of 80\,K/hour followed by quenching in water. The quenched ingots were further ground using mortar-pestle to obtain a homogeneous mixture. The spark plasma sintering (SPS) technique was used for compacting the powders using a graphite dye of 12\,mm in diameter. The sintering was done at 823\,K in Ar atmosphere with a heating and cooling rate of 100 K/minute and 50 K/minute respectively with a hold time of 15 minutes at 823 K under uniaxial pressure of 60 MPa. The pressure was released slowly while cooling to avoid any crack in the sample due to thermal expansion. The sintered pellets were cut to the proper dimensions for electrical and thermal transport measurements. The structural characterization was done using the X-ray diffraction technique using Bruker instrument ($\lambda$=1.5406$\AA$). The surface morphology and chemical composition were observed on the surface of the polished (using an automatic grinding/polishing machine) pellet employing a scanning electron microscope (SEM) equipped with energy dispersive x-ray spectroscopy (EDS) technique, respectively. The Seebeck coefficient ($\alpha$) and electrical conductivity ($\sigma$) was measured using four probe configuration under the Ar atmosphere over a wide temperature range of 300\,K-773\,K. The total thermal conductivity ($\kappa$) was calculated using the following relation: $\kappa$=D$\rho C_p$. The thermal diffusivity (D) was measured using the NETSCH instrument under Ar atmosphere, $\rho$ is the density of the sample calculated using the sample mass and its geometric volume, $C_p$ is the specific heat capacity calculated using Dulong-Petit's law. The uncertainty in the measurement of $\alpha$ and $\sigma$ was 7\% and 5\%, respectively; the estimated uncertainty in $D$ is 5\%. The carrier concentration ($n$) of the samples was measured by Hall measurement at 300\,K and applied magnetic field changing gradually between -1.0 T and 1.0 T, using a homemade setup.
\begin{figure*}
\centering
\includegraphics[width=0.85\linewidth]{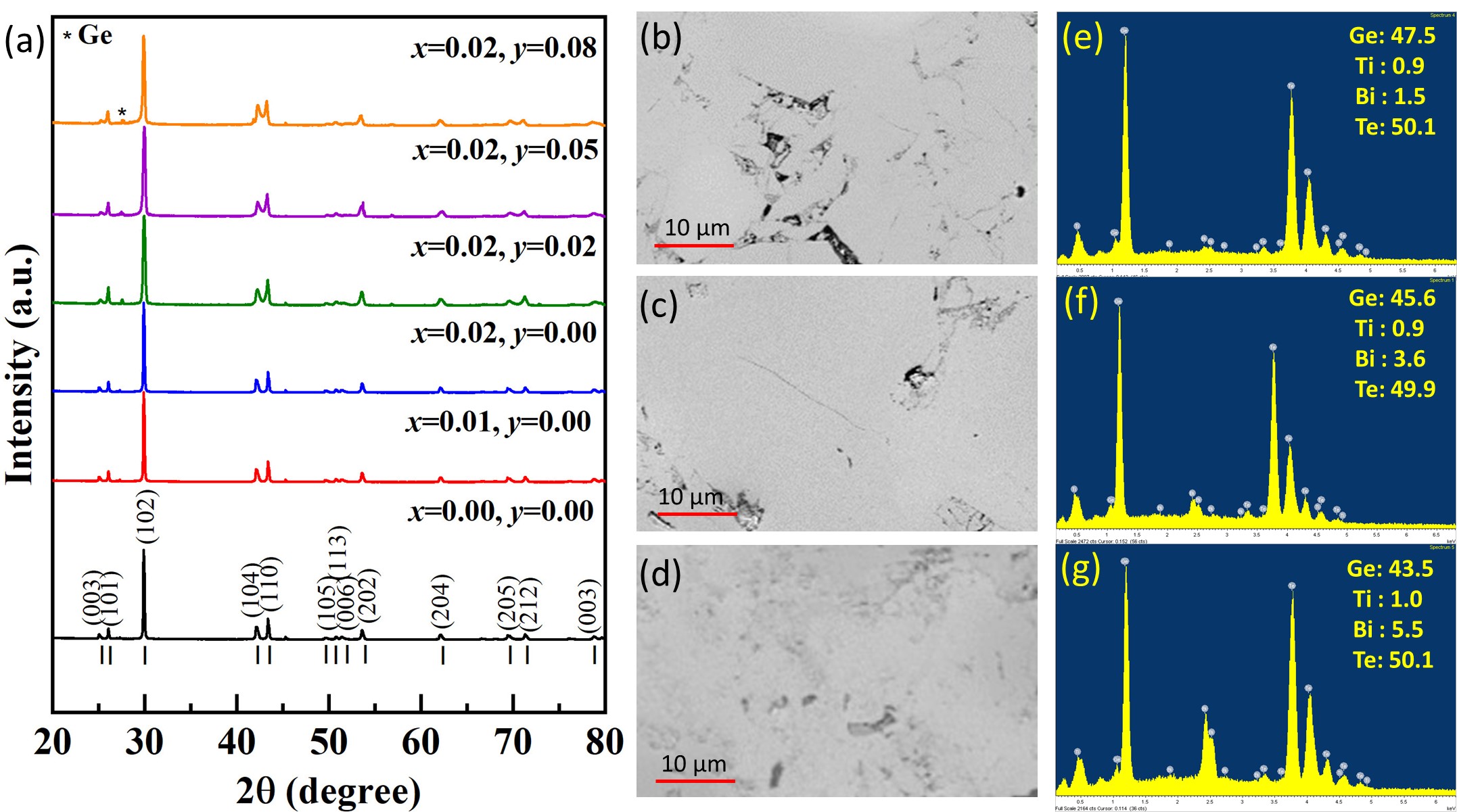}
\caption{(a) X-ray diffraction (XRD) pattern of Ge$_{1.01-x-y}$Ti$_x$Bi$_y$Te (0$\leq$\textit{x}$\leq$0.02, 0$\leq$\textit{y}$\leq$0.08) at 300\,K. The Bragg's position and corresponding Miller indices are marked for GeTe. Scanning electron microscopy image and corresponding energy dispersive x-ray spectra (EDS) for Ge$_{1.01-x-y}$Ti$_x$Bi$_y$Te (b,e) \textit{x}=0.02, \textit{y}=0.02, (c,f) \textit{x}=0.02, \textit{y}=0.05, and (d,g) \textit{x}=0.02, \textit{y}=0.08 is shown.}
\label{XRD}
\end{figure*}
\subsection{Computational Details}
The density functional theory (DFT)\cite{hohenberg1964inhomogeneous, kohn1965self} calculations were carried out using the plane-wave-based pseudopotential approach, as implemented in the Vienna Ab initio Simulation Package (VASP)\cite{kresse1996efficiency, kresse1999ultrasoft}. The structural optimization of all the modeled structures was performed using generalized gradient approximation (GGA) expressed by the Perdew–Burke–Ernzerhof (PBE)\cite{perdew1996generalized} exchange-correlation ($\epsilon_{xc}$) functional. The self-consistency loop was converged with a total energy threshold of 0.01 meV by conjugate gradient (CG) minimization. The structures were fully relaxed until the Heymann–Feynman forces on each atom were less than 10$^{–5}$ eV/Å for both pure and doped configurations. The effects of doping were considered by substituting Ti and Bi atoms at the specific sites of Ge atoms in a 2×2×2 supercell consisting of 48 atoms. All the structures were visualized through VESTA (Visualization for Electronic and STructural Analysis)\cite{momma2011vesta} software. Spin–orbit coupling (SOC) interactions owing to heavy atoms were included when calculating the electronic band structures and density of states. A 6×6×2 $k$-mesh was used for Brillouin zone sampling. The electron wave function was expanded in a plane-wave basis set with an energy cutoff of 600 eV. Phonon calculations were obtained within the harmonic approximation and using a finite displacement method\cite{parlinski1997first}. The phonon calculations are performed using the PHONOPY package\cite{togo2015first, togo2008first}. A 2×2×2 supercell was set for the cubic GeTe containing 64 atoms, whereas, for the rhombohedral phase, a 3×3×1 supercell containing 54 atoms was built. In the Ti–Bi co-doped rhombohedral system, we used a 2×2×2 supercell consisting of 96 atoms. The BoltzTrap Code\cite{madsen2006boltztrap}, based on Boltzmann transport theory, is used to evaluate thermoelectric properties. The starting parameters for the calculations were the values obtained from the refinement.
\section{Results and Discussion}
\subsection{Structural Analysis}
The X-ray diffraction (XRD) pattern for Ge$_{1.01-x-y}$Ti$_{x}$Bi$_{y}$Te; (0$\leq$\textit{x}$\leq$0.02, 0$\leq$\textit{y}$\leq$0.08) samples after spark plasma sintering (SPS) are shown in FIG.~\ref{XRD}(a). The XRD pattern reveals a single phase formation of all the samples with minor peaks corresponding to Ge (shown by $*$ in FIG.~\ref{XRD}(a)). This is attributed to the stoichiometric Ge-rich content present in each sample. The XRD patterns are further analyzed with rhombohedral structure (space group:\textit{R3m}) using Rietveld refinement employing Fullprof software. The refinement pattern for Ge$_{0.91}$Ti$_{0.02}$Bi$_{0.08}$Te is shown in Fig.~S1. The lattice parameters obtained from the Rietveld refinement of the XRD patterns for all the samples are shown in Table~\ref{Table I}. The lattice parameter for $c$-axis decreases from $c$=10.6762 $\AA$ for Ge$_{1.01}$Te to 10.6718 $\AA$ for Ge$_{0.99}$Ti$_{0.02}$Te, whereas the lattice parameter corresponding to $a$-axis increases from $a$=4.1624$\AA$ to 4.1627 $\AA$. The overall $c/a$ ratio and hence the volume of the unit cell decreases with Ti doping and is similar to Mn-doped GeTe.\cite{R25} Bi--doping in Ge$_{0.99-y}$Ti$_{0.02}$Bi$_y$Te further reduces the $c/a$ ratio as observed for only Ti-doped Ge$_{1.01}$Te.\cite{R23} The lattice expansion may be attributed to the reduced Ge vacancies and the larger ionic radii of Ti$^{2+}$ (0.86 $\AA$) and Bi$^{3+}$ (0.96 $\AA$) in contrast to Ge$^{2+}$ (0.73 $\AA$).\cite{shannon}\\
Further surface morphology for Ge$_{1.01-x-y}$Ti$_{x}$Bi$_{y}$Te; with different Ti--Bi co-doping is shown in FIG.\ref{XRD}(b-d) and their corresponding energy dispersive x-ray spectra are shown in FIG.\ref{XRD}(e-g). The homogenous nature of the sample is seen on the surface. The EDS spectra obtained from the surface is shown in FIG.\ref{XRD}(e-g). The atomic \% of the elements present in each sample are shown in the inset of their EDS spectra. The change in atomic \% of the elements is in line with the stoichiometric amount of Ti and Bi in Ge$_{1.01}$Te.
\begin{table}[h]
\caption{Lattice parameters ($a$, $c$), $c/a$, carrier concentration ($n$), and Seebeck coefficient ($\alpha$) for Ge$_{1.01-x-y}$Ti$_{x}$Bi$_{y}$Te ($0.00 \leq x \leq 0.02$,$0.00 \leq y \leq 0.08$) at 300\,K.}
\centering
\begin{tabular}{c c c c c c}
\hline
sample & a  & c & c/a & $n \times$10$^{20}$ & $\alpha$ \\
   $x$, $y$    &($\AA$) & ($\AA$) &  & (cm$^{-3}$) & ($\mu$V/K) \\
\hline
\hline
0.00, 0.00 & 4.1624 & 10.6762 & 2.5649 & 5.20 &38\\
0.01, 0.00 & 4.1625 & 10.6745 & 2.5644 & 5.16 &43\\
0.02, 0.00 & 4.1627 & 10.6718 & 2.5636 & 5.02 &48\\
0.02, 0.02 & 4.1642 & 10.6502 & 2.5575 & 4.10 &65\\
0.02, 0.05 & 4.1662 & 10.6212 & 2.5493 & 2.71 &90\\
0.02, 0.08 & 4.1696 & 10.5816 & 2.5377 & 1.02 &115\\
\hline
\end{tabular}
\label{Table I}
\end{table}
\begin{figure*}
\centering
\includegraphics[width=0.95\linewidth]{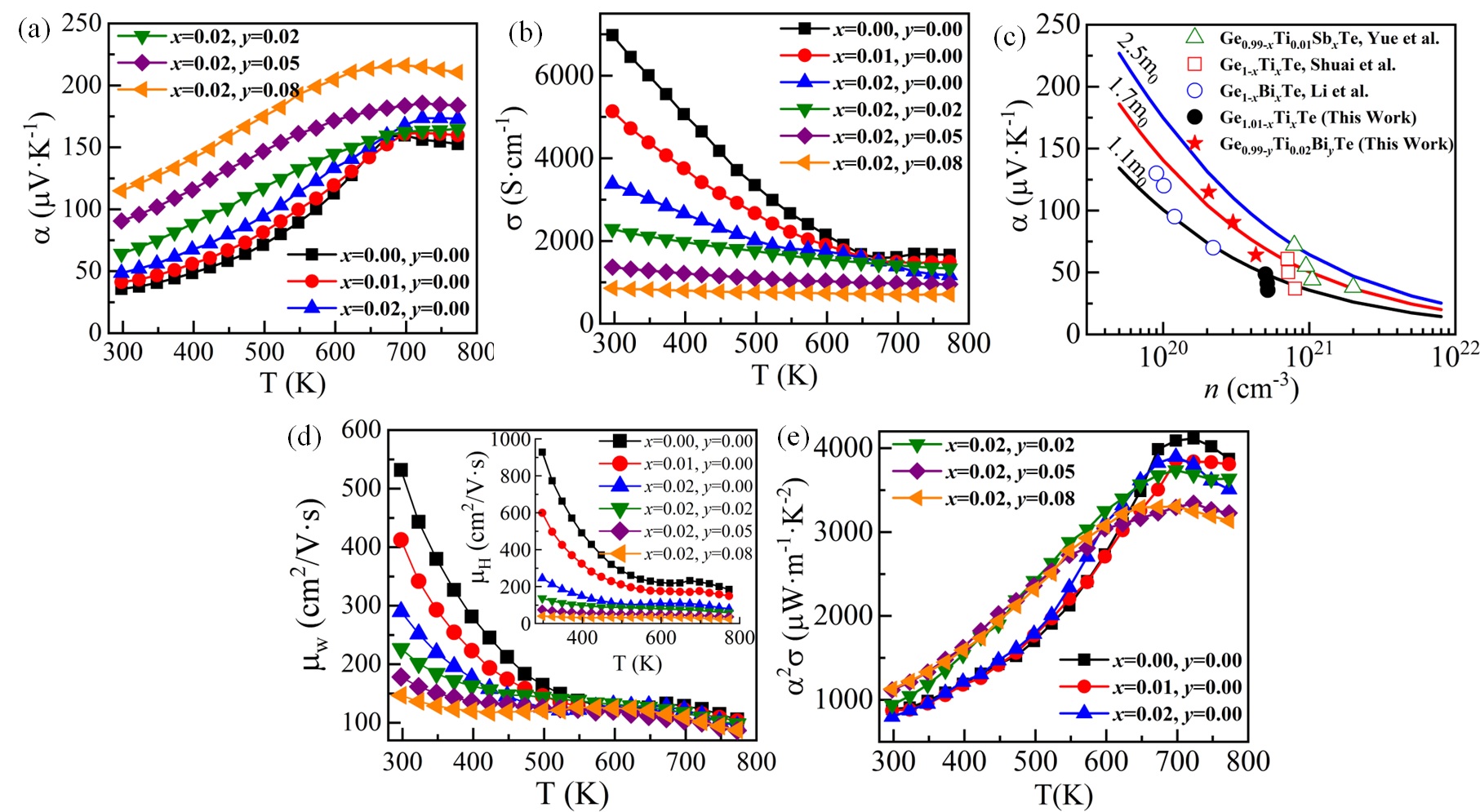}
\caption{Temperature-dependent (a) electrical conductivity ($\sigma$), and (b) Seebeck coefficient ($\alpha$) for Ge$_{1.01-x-y}$Ti$_x$Bi$_y$Te. (c) room-temperature carrier concentration ($n$) dependence of $\alpha$ (Pisarenko plot), compared with literature Yue at. al;\cite{R23}, Shuai et. al;\cite{R27}, Li et. al; \cite{R32}, (d) Temperature-dependent weighted mobility ($\mu_w$), Inset: Hall mobility ($\mu_H$) and (e) temperature-dependent power-factor ($\alpha^2\sigma$) for Ge$_{1.01-x-y}$Ti$_x$Bi$_y$Te (0$\leq$\textit{x}$\leq$0.02, 0$\leq$\textit{y}$\leq$0.08).}
\label{alpha-all}
\end{figure*}
\subsection{Electronic Properties: Experimental}
The temperature-dependent electrical conductivity ($\sigma$) for Ge$_{1.01-x-y}$Ti$_{x}$Bi$_{y}$Te; (0$\leq$\textit{x}$\leq$0.02, 0$\leq$\textit{y}$\leq$0.08) is shown in FIG.~\ref{alpha-all}(a). The $\sigma$ of Ge$_{1.01}$Te is 6970 S·cm$^{-1}$, smaller compared to pristine GeTe ($\sim$ 7800 S·cm$^{-1}$) at 300\,K and is ascribed to decrease in carrier concentration ($n$) in Ge$_{1.01}$Te with diminishing Ge vacancies.\cite{R26} The formation energy of Ge vacancy for Ge rich GeTe phase is larger than that of Te-rich GeTe.\cite{R28} This helps to suppress the Ge vacancy in the GeTe system, and hence a lower carrier concentration is obtained. However, the electrical mobility ($\mu$) improves with reduced $n$ owing to (i) the interaction between the holes decreasing with reduced $n$, and (ii) reduction in Ge vacancies also weakens the scattering of carriers.\cite{R27} It is worth noting that the dimension of holes is closer to the mean free path of the carriers, whereas Ge precipitates are in micron size. Hence the Ge precipitates formed in Ge-rich samples do not influence the $\mu$ as compared to Ge vacancies due to their respective dimensions. In other words, the Ge-rich sample (Ge$_{1.01}$Te) reduces carrier concentration due to a reduction in intrinsic vacancies and enhances $\mu$. The $\sigma$ for Ge$_{1.01}$Te decreases with temperature, indicating a degenerate semiconductor behavior. A rapid change in $\sigma$ across the temperature range of 600-700\,K may be ascribed to the switching of the valence band between $L$ and $\Sigma$ points mediated by the structural transition.\cite{structural} Further, the Ti and Bi doping decreases $\sigma$ across the temperature range studied and it lowers the structural transition temperature as well. The $\sigma$ reduces from 6970 S·cm$^{-1}$ for Ge$_{1.01}$Te to 3350 S·cm$^{-1}$ for Ge$_{0.99}$Ti$_{0.02}$Te to 856 S·cm$^{-1}$ for Ge$_{0.91}$Ti$_{0.02}$Sb$_{0.08}$Te at 300\,K. It is noted that the Ti$^{2+}$ doping does not reduce the $n$ significantly (see Table~\ref{Table I}). Further, the decrease in $\sigma$ with Bi doping is ascribed to a strong reduction in $n$ from 5.02 $\times$ 10$^{20}$ cm$^{-3}$ in Ge$_{0.99}$Ti$_{0.02}$Te to 1.74 $\times$ 10$^{20}$ cm$^{-3}$ in Ge$_{0.91}$Ti$_{0.02}$Sb$_{0.08}$Te at 300\,K. Bi$^{3+}$ doping significantly reduces $n$ by supplying extra valence electrons. Therefore, the carrier concentration is optimized with co-doping of Ti and Bi which matches the optimized carrier concentration.\\
%
%\begin{figure}
%	\centering
%	\includegraphics[width=0.7\linewidth]{geom.jpg}
%	\caption{Crystal structure of (a) Ge$_{23}$Te$_{24}$ (b)  Ge$_{22}$TiTe$_{24}$ (c)  Ge$_{21}$TiBiTe$_{24}$ (d)  Ge$_{20}$TiBi$_2$Te$_{24}$.}
%\end{figure}
%
%
%
The Seebeck coefficient ($\alpha$) as a function of temperature Ge$_{1.01-x-y}$Ti$_{x}$Bi$_{y}$Te ; (0$\leq$\textit{x}$\leq$0.02, 0$\leq$\textit{y}$\leq$0.08) is shown in FIG.~\ref{alpha-all}(b). The positive sign of $\alpha$ indicates a p-type nature of transport and indicates the dominating role of holes in electronic transport. It can be readily seen that the $\alpha$ for Ge$_{1.01}$Te is $\sim$38 $\mu$V·K$^{-1}$ at 300\,K, higher than pristine GeTe ($\sim$ 27 $\mu$V·K$^{-1}$). This rise in $\alpha$ is due to a reduction in $n$ for Ge-rich samples. Further, $\alpha$ increases with Ti doping, since Ti doping may induce resonant levels near the Fermi levels owing to the existence of partly filled spin-up channels. However, due to the high energy of d-orbitals, the resonant level is induced close to the bottom of the conduction band and hence may not act to enlarge $\alpha$. Further, it is noted that Ti doping does not reduce $n$, and the rise in $\alpha$ is attributed to the change in crystal field due to decreased $c/a$ ratio in the sample.\cite{R27} Ti doping does improve $\alpha$, however the $n$ needs to be further reduced. Hence, Bi doping is further induced to optimize $n$ and is also beneficial in further improving $\alpha$ and reducing $\kappa_{ph}$.\cite{R31} As can be seen in Table~\ref{Table I}, the $n$ decreases with Bi doping and results in rise in $\alpha$ from 48 $\mu$V·K$^{-1}$ (Ge$_{0.99}$Ti$_{0.02}$Te) to 115 $\mu$V·K$^{-1}$ (Ge$_{0.91}$Ti$_{0.02}$Sb$_{0.08}$Te) at 300\,K). The $\alpha$ increases with the rise in temperature for all the samples.\\
The change in $\alpha$ with Ti and Bi doping is further analyzed using the Pisarenko plot, as shown in FIG.~\ref{alpha-all}(c). The $\alpha$ dependence on $n$ is determined using the Kane model and two-band approximation (with different effective mass) and assuming the acoustic phonons (\textit{r}=0) as the primary scattering mechanism.\cite{R32}  The $\alpha$ in two-band approximation is expressed as: 
\begin{equation}\label{equation1}
    \alpha=-\frac{k_B}{e}\left[\frac{I_{r+1/2}^1(\eta^*,\beta)}{I_{r+1/2}^0(\eta^*,\beta)}-\eta^*\right]
\end{equation}
where $e$ is electronic charge, $\beta$=$k_BT$/$e_g$, $e_g$ is band gap, $r$ is scattering parameter, $\eta^*$=($E_F-E_V$)/$k_BT$ is the reduced Fermi energy, $I_{i,j}^k(\eta^*,\beta)$ are two parametric Fermi integral given as:\cite{R33, R34}
\begin{equation}\label{equation2}
    I_{i,j}^k(\eta^*,\beta)=\left(-\frac{df}{dx}\right)\frac{x^k(x+\beta x^2)^idx}{(1+2\beta x)^j}
\end{equation}
Further the carrier concentration ($n$), using the obtained values of $\eta^*$, is given as
\begin{equation}\label{equation3}
    n=\frac{(2m_{d,0}k_BT)^{3/2}}{2\pi^2h^3}I_{3/2,0}^0(\eta^*, \beta)
\end{equation}
where $m_{d,0}$=$N_V^{2/3}(m_n^*)^{1/3}(m_{\perp}^*)^{2/3}$ is the band-edge density of states effective mass, $N_V$ is band degeneracy.\\ 
The $n$ does not improve with Ti doping as in the case of Mn doping, however, the effective mass increases with an increase in the Ti content in the system.\cite{R26} Further, Ti--Bi co-doping reduces $n$ and gives rise to $\alpha$. The effective mass increases in the Ti--Bi co-doped samples. In the present case, the effective mass increases from 1.10 (Ge$_{0.99}$Ti$_{0.02}$Te) to 2.13 for Ge$_{0.91}$Ti$_{0.02}$Bi$_{0.08}$Te. This indicates that Bi doping enhances the effective mass due to dominating conduction of the heavy hole valence band over the light hole valence band. However, it is worth noting that the value of effective mass depends on the choice of approximation used for the calculations (such as single parabolic band, Kane model) as well as band gap values, offset between the light and heavy hole bands, etc and hence comparison with literature may not be precise. The increase in $\alpha$ for Ti--Bi co-doped sample is ascribed to the simultaneous optimization of $n$ (through excess Ge content and Bi doping), and band engineering (band degeneration due to Bi and Ti co-doping).\\
An understanding of charge carrier mobility is important for engineering semiconductor materials, which can be measured using the Hall effect. However, the weighted carrier mobility ($\mu_w$), which is defined as the carrier mobility weighted by the density of electronic states, can be estimated using the measured values of the Seebeck coefficient and electrical resistivity using the following equation, as suggested by Synder et al.\cite{mobility}
\begin{multline}
    \mu_w=\frac{3h^3\sigma}{8 \pi e(2m_ek_BT)^{3/2}} \times\\ \left[\frac{exp\left[\frac{|\alpha|}{k_B/e}-2\right]}{1+exp\left[-5\left(\frac{|\alpha|}{k_B/e}-1\right)\right]}+   \frac{\frac{3}{\pi^2}\frac{|\alpha|}{k_B/e}}{1+exp\left[5\left(\frac{|\alpha|}{k_B/e}-1\right)\right]}\right]
\end{multline}
where $k_B$ is Boltzmann constant, $m_e$ is mass of an electron, $h$ is Planck constant, $e$ is electronic charge. Also, $\mu_w$ is related to Hall mobility ($\mu_H$) by the relation
\begin{equation}\label{equation5}
    \mu_w\approx \mu_H \left(\frac{m^*}{m_e}\right)
\end{equation}
with $m^*$ as the density of states electronic mass (effective mass). Since the density of electronic states is proportional to $m^{*3/2}$, $\mu_w$ is considered as electron mobility weighted by the density of states. The $\mu_w$ obtained for Ge$_{1.01-x-y}$Ti$_{x}$Bi$_{y}$Te ; (0$\leq$\textit{x}$\leq$0.02, 0$\leq$\textit{y}$\leq$0.08) as a function of temperature is shown in FIG.\ref{alpha-all}(d). The inset shows the Hall mobility obtained from the $\mu_w$ and corresponding effective mass using equation 5. The change in $\mu_w$ for GeTe indicates a clear acoustic phonon-dominated transport, however, the samples with Ti--Bi co-doping depict a weaker trend. Furthermore, the rise in $\mu_w$ at around 600-650\,K may be ascribed to the rhombohedral–cubic structural phase transition, that leads to the convergence of $\Sigma$ and $L$ band and hence indicates a rise in the Seebeck coefficient. Also, the decrease in $\mu_w$ with temperature is indicative of enhanced scattering of carriers by phonons and holes.\\
The temperature-dependent $\alpha$ and $\sigma$ are used to calculate the power factor ($\alpha^2\sigma$) for Ge$_{1.01-x-y}$Ti$_{x}$Bi$_{y}$Te ; (0$\leq$\textit{x}$\leq$0.02, 0$\leq$\textit{y}$\leq$0.08) system and is shown in FIG.~\ref{alpha-all}(e). The $\alpha^2\sigma$ increases with temperature for all the samples,  however, it tends to decrease at higher temperatures possibly due to phase transition. The power factor for Ti--Bi co-doped samples are larger at a lower temperature due to a significant rise in $\alpha$.
\subsection{Electronic Properties: Theoretical Insights}
\begin{figure*}
\centering
\includegraphics[width=0.88\linewidth]{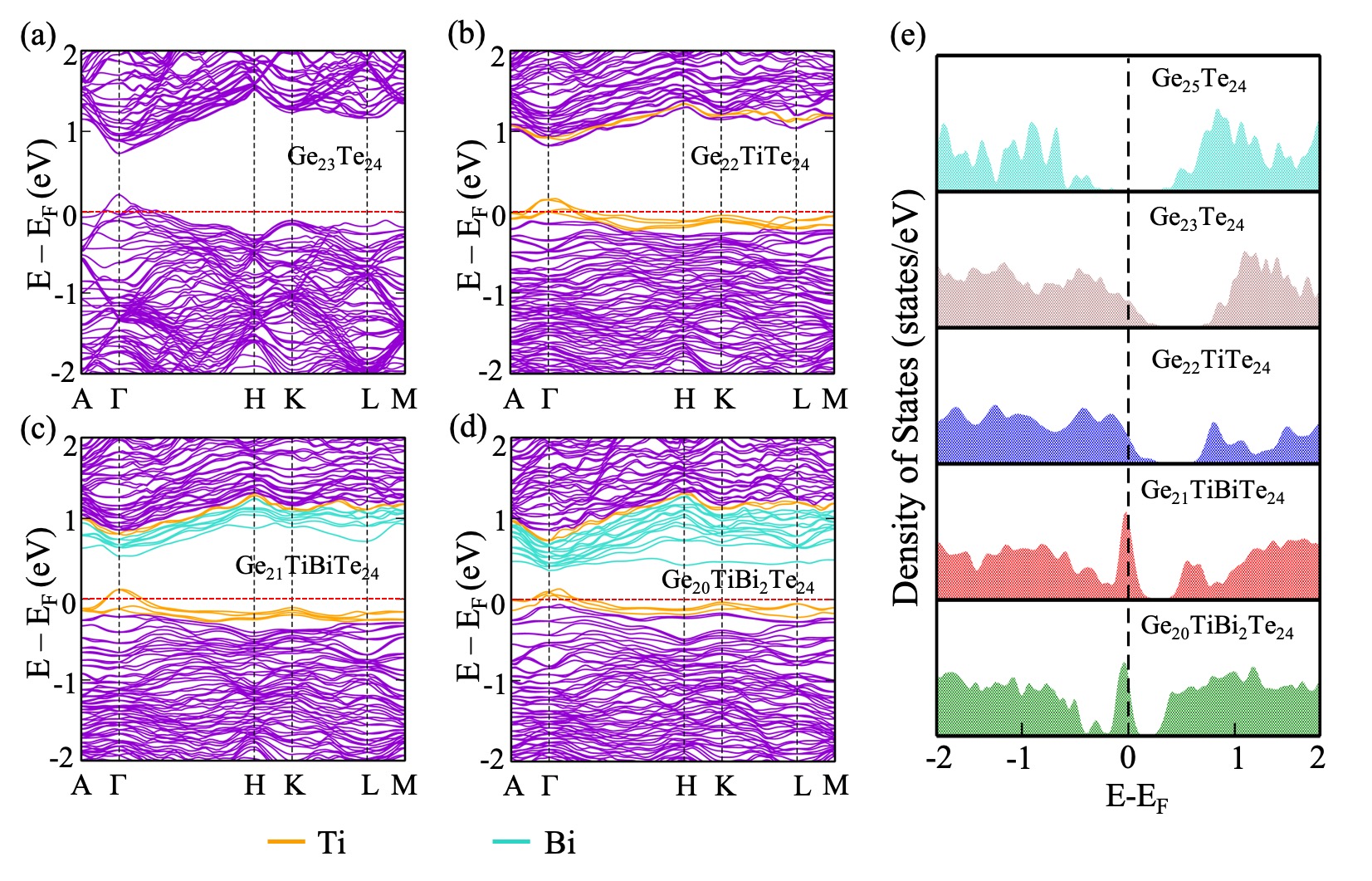}
\caption{Electronic band structures of (a) Ge$_{23}$Te$_{24}$, (b) Ge$_{22}$TiTe$_{24}$, (c) Ge$_{21}$TiBiTe$_{24}$, and (d) Ge$_{20}$TiBi$_2$Te$_{24}$. The Ge vacancies are theoretically induced during the calculation to attest to the large experimental value of carrier concentration. The band gap appears at the $\Gamma$ point in a 2 $\times$ 2 $\times$ 2 supercell containing 48 atoms. The VBM and CBM occur at the L point in the pristine GeTe fold onto the $\Gamma$ point in the supercell. (e) Density of states of GeTe with excess Ge (Ge$_{25}$Te$_{24}$), GeTe with Ge vacancies (Ge$_{23}$Te$_{24}$), Ti-doped GeTe (Ge$_{22}$TiTe$_{24}$), and Ti--Bi co-doped GeTe (Ge$_{21}$TiBiTe$_{24}$, Ge$_{20}$TiBi$_2$Te$_{24}$) samples.}
\label{band}
\end{figure*}
\begin{figure*}
	\centering
	\includegraphics[width=0.75\linewidth]{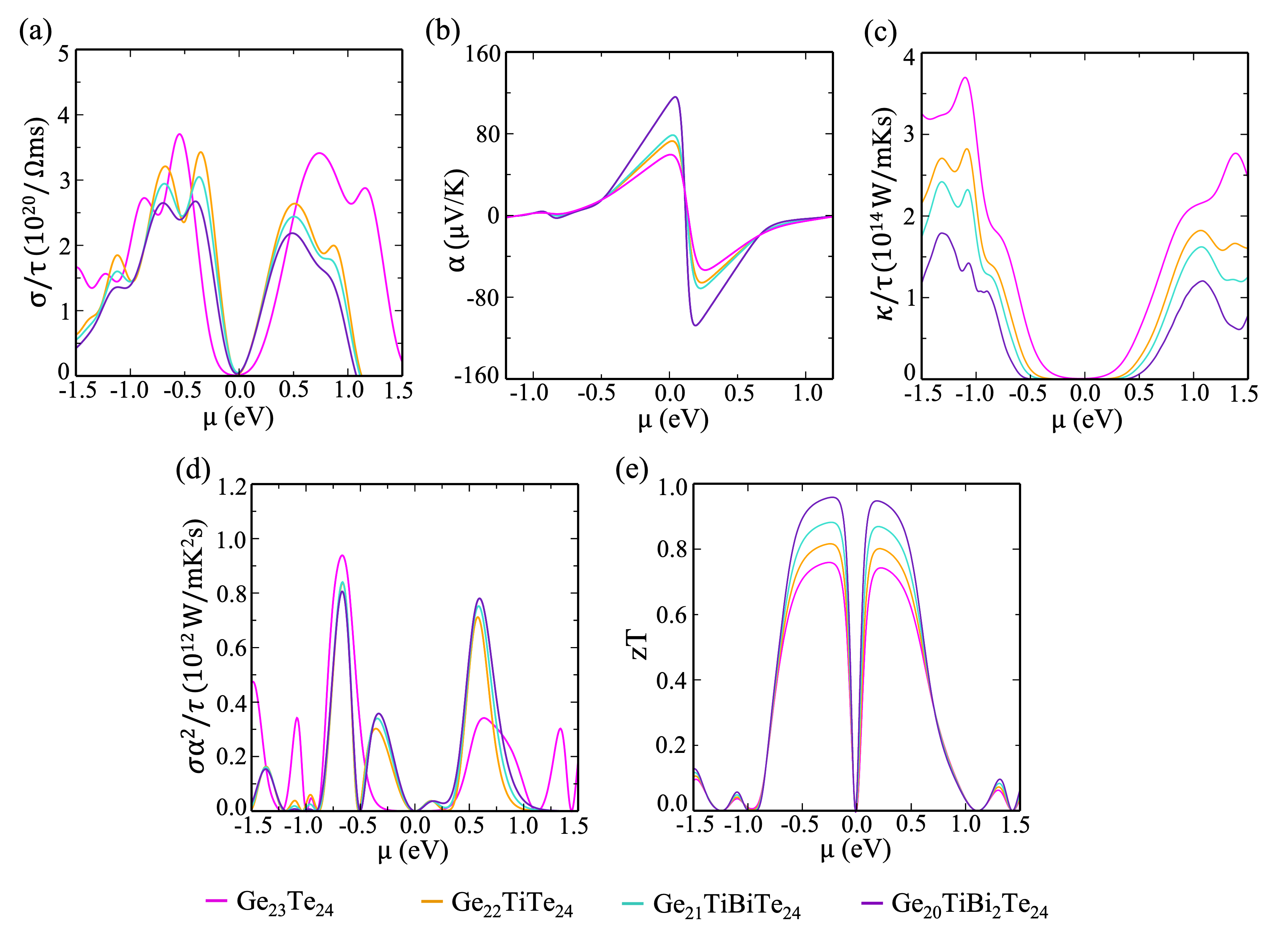}
	\caption{(a) Electrical conductivity ($\sigma$), (b) Seebeck coefficient ($\alpha$), (c) thermal conductivity ($\kappa$), (d) power factor ($\alpha^2\sigma$), and (e) figure of merit (zT) as a function of chemical potential ($\mu$) are calculated at 300\,K for different GeTe compositions: Ge$_{23}$Te$_{24}$, Ge$_{22}$TiTe$_{24}$, Ge$_{21}$TiBiTe$_{24}$ and Ge$_{20}$TiBi$_2$Te$_{24}$. Electrical conductivity, Seebeck coefficient, thermal conductivity, power factor, and figure of merit are reported by scaling them with $\tau$.}
	\label{thermo}
\end{figure*}
To gain insights into the experimental observations regarding the electronic transport in the Ti--Bi co-doped GeTe system, the DFT calculations were carried out to determine electronic band structures and density of states for pristine, Ti-doped, and Ti--Bi co-doped GeTe. We have systematically doped Ge, Ti, and Ti--Bi in the GeTe system to obtain all the required configurations. The present calculations show that the principal valence band (light hole) maximum (VBM) and conduction band minimum (CBM) occur at the $\Gamma$ point due to the folding of the L point onto $\Gamma$ (see Fig.~S2 in SI). As the pristine GeTe shows a high hole carrier concentration owing to the intrinsic Ge vacancies present in it, the same has been considered in DFT calculations. The electronic band structure for Ge$_{23}$Te$_{24}$ is shown in FIG.~\ref{band}(a). Further, (i) one Ti (FIG.~\ref{band}(b) (ii) one Ti and one Bi (FIG.~\ref{band}(c)) (iii) one Ti and two Bi atoms in place of Ge were substituted in the GeTe supercell. For the Ti--Bi co-doped case, the energies of the two configurations were calculated, showing that (i) Ti and Bi atoms are close to each other and (ii) they are far from each other. The latter configuration has lower energy, making it more stable, and thus the second configuration is considered for all further calculations. As we can see in FIG.~\ref{band}b, the new impurity bands arise from the Ti states near the conduction band, which reduces the band gap. These resonant levels are induced below the conduction band owing to the high energy Ti-$d$ orbitals and are consistent with the previous studies.\cite{R27} The co-doping of Ti and Bi at Ge in GeTe (Ge$_{21}$TiBiTe$_{24}$) further decreases the energy separation between the valence bands, $\Delta$E$_\Gamma$, from 0.21 to 0.13 eV, which is further decreased to 0.08 eV in Ge$_{20}$TiBi$_2$Te$_{24}$, leading to band convergence in GeTe. This indicates that the heavy holes strongly influence carrier transport. Therefore, the Ti--Bi co-doping in GeTe enhances the valence band convergence and confirms the obtained increase in the Seebeck coefficient.\\
Further, we have plotted the density of states (DOS) of Ge$_{25}$Te$_{24}$ (with excess Ge), followed by Ge$_{23}$Te$_{24}$ (with Ge vacancies), Ge$_{22}$TiTe$_{24}$ (Ti-doped) and Ge$_{21}$TiBiTe$_{24}$ and Ge$_{20}$TiBi$_2$Te$_{24}$ (Ti--Bi co-doped) samples. The Fermi level position plays a vital role in optimizing the TE performance of any system. The high hole carrier concentration in GeTe deepens the Fermi level in the valence band.  However, the density of states (DOS) plot shows that the Fermi level lies in the middle of the band gap for pristine Ge$_{24}$Te$_{24}$\cite{R26}. Therefore, we have theoretically introduced Ge vacancies in GeTe, due to which the Fermi level lies deep in the valence band and is in line with the experimental high carrier concentration present in GeTe. With excess Ge (Ge$_{25}$Te$_{24}$), the Fermi level tends to move towards the conduction band. The Ti doping in GeTe (with excess Ge) increases the carrier concentration, resulting in the shifting of E$_F$ deeper in the valence band. Bi doping reduces the hole carrier concentration due to the donar nature of Bi. The decrease in the energy offset between the valence band edges results in band convergence in the co-doped system, as shown in band structure calculations. Therefore, the combination of both band convergence and the Fermi level position indicates an enlarged Seebeck coefficient in co-doped samples. In addition, Ti--Bi co-doping makes the DOS steeper, particularly near the valence band edge. This sharper DOS feature indicates a higher effective mass and enhances the Seebeck coefficient\cite{liu2018phase}. The Fermi level lies in the valence band for both Ti and Bi co-doping because of the larger Ge vacancies considered during theoretical calculations.\\
%\\
%
%
Subsequently, to examine the thermoelectric properties, we have calculated the electrical conductivity ($\sigma$), Seebeck coefficient ($\alpha$), thermal conductivity ($\kappa$), power factor ($\alpha^2\sigma$), and thermoelectric figure of merit ($zT$) for Ge$_{23}$Te$_{24}$, Ge$_{22}$TiTe$_{24}$, Ge$_{21}$TiBiTe$_{24}$ and Ge$_{20}$TiBi$_2$Te$_{24}$ systems as a function of chemical potential ($\mu$). Note that all $\mu$ dependent transport properties are calculated by fixing the temperature and letting carrier concentration ($n$) change. It implies that for a fixed temperature (here 300\:K), the $\mu$ is a function of $n$. The study of thermoelectric properties with $\mu$ interprets the doping of carriers as it shows the addition/removal of electrons in/from the system. % Note that all $\mu$ dependent transport properties are calculated by fixing the temperature and letting carrier concentration ($n$) change. It implies that for a fixed temperature (here 300\:K), the $\mu$ is a function of $n$.   
Advantageously, the position of $\mu$ determines the fraction of electrons in the conduction or valence band which take part in the electronic transport and hence influences the transport properties. Thus, we determine the thermoelectric parameters as a function of $\mu$. FIG.~\ref{thermo}(b) shows the variation of the Seebeck coefficient with $\mu$.  $\alpha$ measures the induced thermoelectric voltage ($\Delta$V) in response to a temperature difference ($\Delta$T) in the material and is given as $\alpha$=$\Delta$V/$\Delta$T. By definition, $\mu$=0 coincides with the top of the valence band in semiconductors. This implies that at $\mu$=0, the nature of $\alpha$ determines the type of semiconductor. From FIG.~\ref{thermo}(b), we can see that at $\mu$=0, the value of $\alpha$ is positive for all GeTe systems, indicating that these are $p$-type semiconductors. The $\alpha$ is plotted as a function of temperature for the GeTe systems (see Fig. S4 in SI). From FIG.~\ref{thermo}a and ~\ref{thermo}c, we see that electrical and thermal conductivity decreases with Ti--Bi doping, which is in line with the experimental findings. The increase in $\alpha$ with doping results in an increase in power factor values compared to the pristine system (see FIG.~\ref{thermo}(d)). The resonant peaks are observed near the Fermi level in the positive `$\mu$' region. In addition, the zT values (FIG.~\ref{thermo}(e)) are higher in the negative `$\mu$' region than the positive one, indicating that $p$-type doping has higher zT values. This predicts that the considered co-doped systems are promising $p$-type thermoelectric materials.\\
\begin{figure*}
\centering
\includegraphics[width=0.75\linewidth]{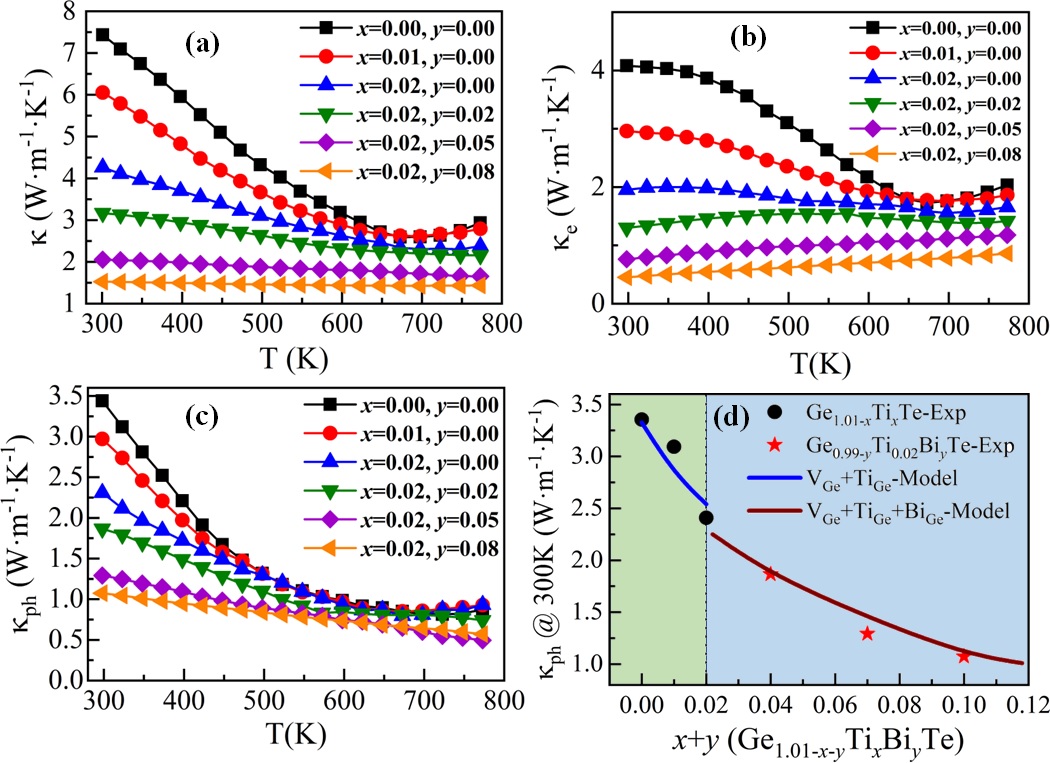}
\caption{(a) Total thermal conductivity ($\kappa$), (b) electronic thermal conductivity ($\kappa_e$), and (c) phonon thermal conductivity ($\kappa_{ph}$) as a function of temperature for Ge$_{1.01-x-y}$Ti$_x$Bi$_y$Te; (0$\leq$\textit{x}$\leq$0.02, 0$\leq$\textit{y}$\leq$0.08). (d) Phonon thermal conductivity ($\kappa_{ph}$) as a function of Ti ($x$) and Bi ($y$) doping in Ge$_{1.01-x-y}$Ti$_x$Bi$_y$Te at 300\,K. The $\kappa_{ph}$ calculated using the Debye-Callaway model is shown also shown.}
\label{kappa}
\end{figure*}
%
%\textcolor{blue}{Figure (b), y-axis should be $\alpha$/($\tau$).}\\
%\textcolor{blue}{This section should be elaborated more in terms of underlying physics that influences the change in $\alpha$, and $\sigma$ with Ti--Bi co-doping.}\\
%\textcolor{blue}{Plz. talk about the Seebeck coefficient calculations as a function of chemical potential at a different temperatures, Fig.~S4 in SI.}\\
%
%
%
%
\subsection{Thermal conductivity}
The total thermal conductivity ($\kappa$) as a function of temperature for Ge$_{1.01-x-y}$Ti$_{x}$Bi$_{y}$Te ; (0$\leq$\textit{x}$\leq$0.02, 0$\leq$\textit{y}$\leq$0.08) is shown in FIG.~\ref{kappa}(a). $\kappa$ for Ti-doped and Ti-Bi co-doped samples is obviously lower than that of Ge$_{1.01}$Te and GeTe systems. $\kappa$ first decreases with Ti-doping, and when Ti-doping is kept constant at $x$=0.02, $\kappa$ decreases with further Bi-doping. Ge$_{0.91}$Ti$_{0.02}$Bi$_{0.08}$Te presents the lower $\kappa$ among all the samples studied across the whole temperature range. Also, $\kappa$ reduces with the rise in temperature for all the samples as often observed in typical degenerate semiconductor solids. The $\kappa$ at 300\,K for pristine Ge$_{1.01}$Te is 7.5 W·m$^{-1}$·K$^{-1}$ and decreases to 2.59 W·m$^{-1}$·K$^{-1}$ at 673\,K and increases with a further rise in temperature due to second-order structural transition and is consistent with the change in $\sigma$ with temperature (FIG.~\ref{alpha-all}(a)). Further, Ti doping lowers $\kappa$ to 4.30 W·m$^{-1}$·K$^{-1}$ at 300\,K for Ge$_{0.99}$Ti$_{0.02}$Te and is prominently due to reduced $\sigma$. Also, the co-doping of Ti and Bi reduces $\kappa$ to 1.53 W·m$^{-1}$·K$^{-1}$ for Ge$_{0.91}$Ti$_{0.02}$Bi$_{0.08}$Te at 300\,K.\\
\begin{figure*}
\centering
\includegraphics[width=0.8\linewidth]{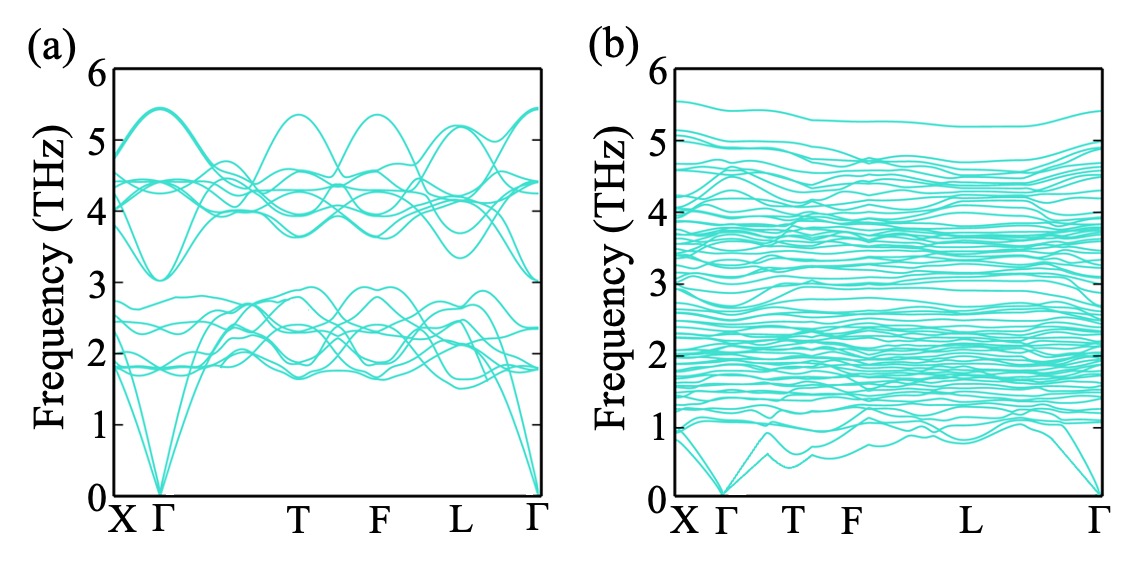}
\caption{Phonon dispersion curve for (c) Ge$_{24}$Te$_{24}$ and (b) Ge$_{21}$TiBi$_2$Te$_{24}$ in a rhombohedral structure.}
\label{phonon}
\end{figure*}
Since $\kappa$ consists of electronic thermal conductivity ($\kappa_e$) and phonon thermal conductivity ($\kappa_{ph}$). It is important to understand the change in $\kappa_e$ and $\kappa_{ph}$ with Ti and Bi co-doping in Ge$_{1.01}$Te. The $\kappa_e$ is calculated using Wiedemann Franz law: $\kappa_e$=$L\sigma$T, where the temperature-dependent Lorenz number ($L$) is obtained by estimating the reduced chemical potential from temperature-dependent Seebeck coefficient considering two-band model.\cite{R26} Temperature-dependent $\kappa_e$ is shown in FIG.~\ref{kappa}(b). The dramatic reduction in $\kappa_e$ is correlated with the reduced $\sigma$ owing to reduced $n$ in co-doped samples. Further, $\kappa_{ph}$ is extracted from $\kappa$ via subtracting $\kappa_e$. The temperature-dependent $\kappa_{ph}$ for Ge$_{1.01-x-y}$Ti$_{x}$Bi$_{y}$Te; (0$\leq$ \textit{x}$\leq$0.02, 0$\leq$\textit{y}$\leq$0.08) is shown in FIG.~\ref{kappa}(c). The $\kappa_{ph}$ for Ge$_{1.01}$Te is 3.43 W·m$^{-1}$·K$^{-1}$ at 300\,K and it reduces to 0.82 W·m$^{-1}$·K$^{-1}$ at 673\,K and then starts to increase above transition temperature. It is possibly due to the ferroelectric instability close to the phase transition temperature as it induces soft optical phonon modes that strongly scatter the heat-carrying phonons.\cite{phonon} Further, it is noted that the $\kappa_{ph}$ for Ge$_{1.01}$Te is also influenced by the reduction in hole concentration. The decrease in hole concentration not only reduces $\sigma$ but also reduces the phonon scattering due to a reduction in $n$ and hence $\kappa_{ph}$ is higher than that of GeTe ($\sim$3 W·m$^{-1}$·K$^{-1}$ at 300\,K). However, with Ti doping the $\kappa_{ph}$ reduces to 2.30 W·m$^{-1}$·K$^{-1}$ and for Ti--Bi co-doped system, a reduced $\kappa_{ph}$ of 1.05 W·m$^{-1}$·K$^{-1}$ is obtained at 300\,K, which further decreases to 0.57 W·m$^{-1}$·K$^{-1}$ at 773\,K for Ge$_{0.91}$Ti$_{0.02}$Bi$_{0.08}$Te. This suggests that Ti and Bi co-doping significantly lowers $\kappa_{ph}$ across the temperature range. FIG.~\ref{kappa}(d) shows the $\kappa_{ph}$ as a function of Ti and Bi co-doping in Ge$_{1.01}$Te at 300\,K. The mass and strain field fluctuation in solid solutions result in point defect scattering. The $\kappa_{ph}$ was calculated using the Debye-Callaway model where the ratio of phonon thermal conductivity with point defects $\kappa_{ph}$ to the parent system $\kappa_{ph}^p$, above the Debye temperature ($\Theta_D$) is expressed as $\frac{\kappa_{ph}}{\kappa_{ph}^p}=\frac{tan^{-1}(u)}{u}$. where $u=\left(\frac{\pi^2 \Theta_D\Omega}{h\nu_a^2}\kappa_{ph}^p\Gamma\right)^{1/2}$ where $\Omega$, $h$ and $\nu_a$ are average volume per atom, Planck constant, and the average value of sound velocity respectively and $\Gamma$ is the scattering parameter consists of mass and strain field fluctuation.\cite{R26} The $\kappa_{ph}$ obtained from the Debye-Callaway model is shown by solid-line in FIG.\ref{kappa}(d) and is in agreement to the $\kappa_{ph}$ obtained from $\kappa$. This decrease in $\kappa_{ph}$ in Ti--Bi co-doped sample is hence attributed to point defect scattering owing to mass and strain field fluctuations and agrees with the previous report of Mn-doped GeTe.\cite{R26}
\subsection{Phonon Dispersion Calculation}
The underlying reason for the reduction in lattice thermal conductivity in Ti--Bi co-doped GeTe can be understood by plotting the phonon dispersion curve. The phonon dispersion calculation for GeTe (Ge$_{24}$Te$_{24}$) and Ti--Bi co-doped (Ge$_{21}$TiBi$_2$Te$_{24}$) [vacancies are not considered while calculating the phonon dispersion plots due to the technical challenges in accommodating so many defects inside a supercell\cite{R26}] was performed to examine the dynamical stability. FIG.~\ref{phonon}(a-b) shows the phonon dispersion plot for Ge$_{24}$Te$_{24}$  and Ge$_{21}$TiBi$_2$Te$_{24}$ in the rhombohedral structure. The phonon dispersion curve for GeTe is consistent with previous studies\cite{wdowik2014soft}. The phonon dispersion curve for Ti--Bi co-doped shows a significant decrease in the phonon dispersion curve slope, as shown in FIG.~\ref{phonon}(b). Generally, the phonon dispersion is shown by the $\omega$ vs $k$ plot, and the gradient of $\omega$ vs $k$ curve measures the $v_\textrm{g}$ (phonon group velocity), where $v_\textrm{g}$ = d$\omega$/d$k$. As shown in Fig~\ref{phonon}(b), the gradient of the phonon curve for the Ti--Bi co-doped system is lower than that of GeTe (Fig~\ref{phonon}(a)). This implies that the mean phonon group velocity decreases for Ti--Bi co-doped sample compared to GeTe, leading to a lower lattice thermal conductivity. This decrease in the slope of the dispersion curve for Bi doping can be attributed to the large atomic mass (M) of Bi, as $\omega$ $\propto$ $M^{-1/2}$. Further, to examine the charge transfer, the charge density contours for Ge$_{24}$Te$_{24}$ and Ge$_{21}$TiBi$_2$Te$_{24}$ are computed (see SI, Fig.~S3). As shown in Fig.~S3, there is a charge transfer between Bi and Ge atoms, which further stabilizes the bond. This stable bond hinders the free vibration of atoms and hence lowers the phonon group velocity.

\subsection{Thermoelectric Figure of merit}
\begin{figure}
\centering
\includegraphics[width=0.8\linewidth]{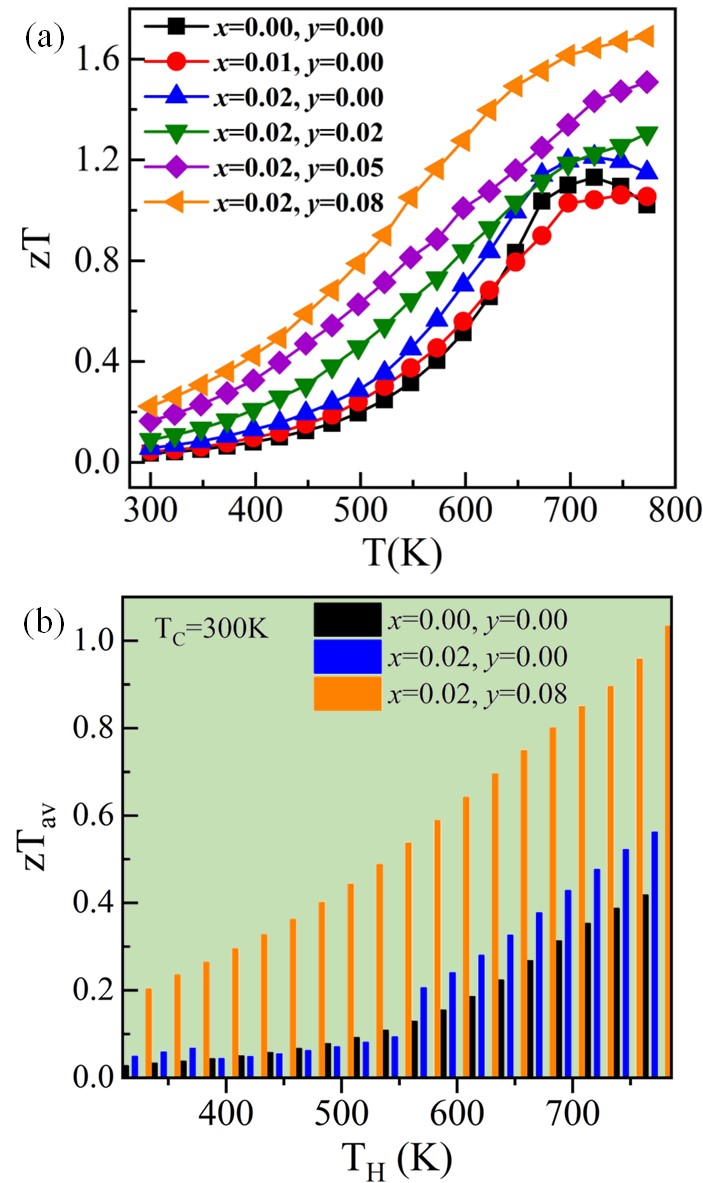}
\caption{(a) Figure of merit (zT), and (b) average figure of merit (zT$_{av}$) as a function of temperature for Ge$_{1.01-x-y}$Ti$_x$Bi$_y$Te ; (0$\leq$\textit{x}$\leq$0.02, 0$\leq$\textit{y}$\leq$0.08).}
\label{zT}
\end{figure}
The TE figure of merit (zT) is calculated using the measured values of $\alpha$, $\sigma$, and $\kappa$ and is shown as a function of temperature for Ge$_{1.01-x-y}$Ti$_{x}$Bi$_{y}$Te ; (0$\leq$\textit{x}$\leq$0.02, 0$\leq$\textit{y}$\leq$0.08) in FIG.~\ref{zT}(a). The zT increases with an increase in temperature for all the samples. A maximum zT of 1.09 is obtained for Ge$_{1.01}$Te at 673\,K, depicting that vacancy engineering in GeTe itself improves its TE performance.\cite{zT1} Further, zT increases to 1.21 at 723\,K in Ti doped samples consistent with the previous studies; \cite{R23} however a significant increase in zT is observed in Ti--Bi co-doped sample. It is attributed to the simultaneous optimization of band structure, carrier concentration, and phonon thermal conductivity. A maximum zT of 1.75 is obtained at 773\,K for Ge$_{0.91}$Ti$_{0.02}$Bi$_{0.08}$Te. The applicability of a TE material is quantified using its average zT (zT$_{av}$=$\int_{T_c}^{T_h}zTdT$). The zT$_{av}$ is calculated using the lower ($T_c$) and upper ($T_h$) limits of temperature as 300\,K and 773\,K respectively, and is shown in FIG.~\ref{zT}(b). A maximum zT$_{av}$ of 1.03 is obtained for Ge$_{0.91}$Ti$_{0.02}$Bi$_{0.08}$Te and which is significantly better as compared to the recently reported several GeTe-based materials.\cite{zT3, zT4, R25, zT6, zT7} The theoretical energy conversion efficiency ($\eta$) is also calculated for Ge$_{0.91}$Ti$_{0.02}$Bi$_{0.08}$Te (p-type leg) and assuming a similar n-type leg for a temperature difference of 473\,K. A maximum $\eta$ of 14\% is obtained in the present study which is higher than several TE materials reported in the literature.\cite{zT3, zT4, R25, zT6, zT7}\\

\section{Conclusion}
This study demonstrates the enhanced thermoelectric properties in Ti--Bi co-doped Ge$_{1.01}$Te accompanied by vacancy engineering. The samples synthesized using the melting-quenching-sintering process show a single-phase formation along with traces of Ge peaks due to its stoichiometric larger content, which is further supported by electron microscopy analysis. The excess Ge reduces the carrier concentration $n$ and hence adjusts the Fermi level position, which improves the TE performance of Ge$_{1.01}$Te. Further, Ti doping improves $\alpha$ due to crystal field engineering ascribed to decreased $c/a$ ratio, which is further enhanced by reducing $n$ with Bi doping. The Ti--Bi co-doped Ge$_{1.01}$Te shows larger band degeneracy and hence improves $\alpha$. The DFT calculations further verify the valence band convergence on Ti--Bi co-doping and
confirm the obtained increase in the $\alpha$. The reduction in total thermal conductivity ($\kappa$) due to decreased electrical conductivity and enhanced point defect and mass fluctuation scattering is obtained. The phonon dispersion calculations show that phonon group velocity reduces in the Ti--Bi co-doped system due to the larger atomic mass of
Bi and the charge transfer between Bi and Ge atoms. A combined effect of carrier concentration optimization, enhanced band degeneracy, and reduced phonon thermal conductivity ($\kappa_{ph}$), a high zT of 1.75 at 773\,K is obtained for Ge$_{0.91}$Ti$_{0.02}$Bi$_{0.08}$Te. This further results in an average zT (zT$_{av}$) of 1.03 for a temperature difference of 473\,K, making it a promising candidate for practical applications. Also, a maximum energy conversion efficiency ($\eta$) of 14\% is calculated assuming similar n-type materials which are promising for thermoelectric devices.

\section{Acknowledgement}
This work was supported by the French Agence Nationale de la Recherche (ANR), through the project NEO (ANR 19-CE30-0030-01). P.B. acknowledges UGC, India, for the senior research fellowship [grant no. 1392/(CSIR-UGC NET JUNE 2018)]. S.B. acknowledges financial support from SERB under a core research grant (Grant No. CRG/2019/000647) to set up his High Performance Computing (HPC) facility “Veena” at IIT Delhi for computational resources.

\newpage

\begin{thebibliography}{13}
\expandafter\ifx\csname natexlab\endcsname\relax\def\natexlab#1{#1}\fi
\expandafter\ifx\csname bibnamefont\endcsname\relax
  \def\bibnamefont#1{#1}\fi
\expandafter\ifx\csname bibfnamefont\endcsname\relax
  \def\bibfnamefont#1{#1}\fi
\expandafter\ifx\csname citenamefont\endcsname\relax
  \def\citenamefont#1{#1}\fi
\expandafter\ifx\csname url\endcsname\relax
  \def\url#1{\texttt{#1}}\fi
\expandafter\ifx\csname urlprefix\endcsname\relax\def\urlprefix{URL }\fi
\providecommand{\bibinfo}[2]{#2}
\providecommand{\eprint}[2][]{\url{#2}}

\bibitem[{\citenamefont{Hohenberg and Kohn}(1964)}]{hohenberg1964inhomogeneous}
\bibinfo{author}{\bibfnamefont{P.}~\bibnamefont{Hohenberg}} \bibnamefont{and}
  \bibinfo{author}{\bibfnamefont{W.}~\bibnamefont{Kohn}},
  \bibinfo{journal}{Physical review} \textbf{\bibinfo{volume}{136}},
  \bibinfo{pages}{B864} (\bibinfo{year}{1964}).

\bibitem[{\citenamefont{Kohn and Sham}(1965)}]{kohn1965self}
\bibinfo{author}{\bibfnamefont{W.}~\bibnamefont{Kohn}} \bibnamefont{and}
  \bibinfo{author}{\bibfnamefont{L.~J.} \bibnamefont{Sham}},
  \bibinfo{journal}{Physical review} \textbf{\bibinfo{volume}{140}},
  \bibinfo{pages}{A1133} (\bibinfo{year}{1965}).

\bibitem[{\citenamefont{Kresse and
  Furthm{\"u}ller}(1996)}]{kresse1996efficiency}
\bibinfo{author}{\bibfnamefont{G.}~\bibnamefont{Kresse}} \bibnamefont{and}
  \bibinfo{author}{\bibfnamefont{J.}~\bibnamefont{Furthm{\"u}ller}},
  \bibinfo{journal}{Computational materials science}
  \textbf{\bibinfo{volume}{6}}, \bibinfo{pages}{15} (\bibinfo{year}{1996}).

\bibitem[{\citenamefont{Kresse and Joubert}(1999)}]{kresse1999ultrasoft}
\bibinfo{author}{\bibfnamefont{G.}~\bibnamefont{Kresse}} \bibnamefont{and}
  \bibinfo{author}{\bibfnamefont{D.}~\bibnamefont{Joubert}},
  \bibinfo{journal}{Physical review b} \textbf{\bibinfo{volume}{59}},
  \bibinfo{pages}{1758} (\bibinfo{year}{1999}).

\bibitem[{\citenamefont{Perdew et~al.}(1996)\citenamefont{Perdew, Burke, and
  Ernzerhof}}]{perdew1996generalized}
\bibinfo{author}{\bibfnamefont{J.~P.} \bibnamefont{Perdew}},
  \bibinfo{author}{\bibfnamefont{K.}~\bibnamefont{Burke}}, \bibnamefont{and}
  \bibinfo{author}{\bibfnamefont{M.}~\bibnamefont{Ernzerhof}},
  \bibinfo{journal}{Physical review letters} \textbf{\bibinfo{volume}{77}},
  \bibinfo{pages}{3865} (\bibinfo{year}{1996}).

\bibitem[{\citenamefont{Momma and Izumi}(2011)}]{momma2011vesta}
\bibinfo{author}{\bibfnamefont{K.}~\bibnamefont{Momma}} \bibnamefont{and}
  \bibinfo{author}{\bibfnamefont{F.}~\bibnamefont{Izumi}},
  \bibinfo{journal}{Journal of applied crystallography}
  \textbf{\bibinfo{volume}{44}}, \bibinfo{pages}{1272} (\bibinfo{year}{2011}).

\bibitem[{\citenamefont{Parlinski et~al.}(1997)\citenamefont{Parlinski, Li, and
  Kawazoe}}]{parlinski1997first}
\bibinfo{author}{\bibfnamefont{K.}~\bibnamefont{Parlinski}},
  \bibinfo{author}{\bibfnamefont{Z.}~\bibnamefont{Li}}, \bibnamefont{and}
  \bibinfo{author}{\bibfnamefont{Y.}~\bibnamefont{Kawazoe}},
  \bibinfo{journal}{Physical Review Letters} \textbf{\bibinfo{volume}{78}},
  \bibinfo{pages}{4063} (\bibinfo{year}{1997}).

\bibitem[{\citenamefont{Togo and Tanaka}(2015)}]{togo2015first}
\bibinfo{author}{\bibfnamefont{A.}~\bibnamefont{Togo}} \bibnamefont{and}
  \bibinfo{author}{\bibfnamefont{I.}~\bibnamefont{Tanaka}},
  \bibinfo{journal}{Scripta Materialia} \textbf{\bibinfo{volume}{108}},
  \bibinfo{pages}{1} (\bibinfo{year}{2015}).

\bibitem[{\citenamefont{Togo et~al.}(2008)\citenamefont{Togo, Oba, and
  Tanaka}}]{togo2008first}
\bibinfo{author}{\bibfnamefont{A.}~\bibnamefont{Togo}},
  \bibinfo{author}{\bibfnamefont{F.}~\bibnamefont{Oba}}, \bibnamefont{and}
  \bibinfo{author}{\bibfnamefont{I.}~\bibnamefont{Tanaka}},
  \bibinfo{journal}{Physical Review B} \textbf{\bibinfo{volume}{78}},
  \bibinfo{pages}{134106} (\bibinfo{year}{2008}).

\bibitem[{\citenamefont{Madsen and Singh}(2006)}]{madsen2006boltztrap}
\bibinfo{author}{\bibfnamefont{G.~K.} \bibnamefont{Madsen}} \bibnamefont{and}
  \bibinfo{author}{\bibfnamefont{D.~J.} \bibnamefont{Singh}},
  \bibinfo{journal}{Computer Physics Communications}
  \textbf{\bibinfo{volume}{175}}, \bibinfo{pages}{67} (\bibinfo{year}{2006}).

\bibitem[{\citenamefont{Kumar et~al.}(2021)\citenamefont{Kumar, Bhumla,
  Parashchuk, Baran, Bhattacharya, and
  Wojciechowski}}]{doi:10.1021/acs.chemmater.1c00331}
\bibinfo{author}{\bibfnamefont{A.}~\bibnamefont{Kumar}},
  \bibinfo{author}{\bibfnamefont{P.}~\bibnamefont{Bhumla}},
  \bibinfo{author}{\bibfnamefont{T.}~\bibnamefont{Parashchuk}},
  \bibinfo{author}{\bibfnamefont{S.}~\bibnamefont{Baran}},
  \bibinfo{author}{\bibfnamefont{S.}~\bibnamefont{Bhattacharya}},
  \bibnamefont{and} \bibinfo{author}{\bibfnamefont{K.~T.}
  \bibnamefont{Wojciechowski}}, \bibinfo{journal}{Chemistry of Materials}
  \textbf{\bibinfo{volume}{33}}, \bibinfo{pages}{3611} (\bibinfo{year}{2021}).

\bibitem[{\citenamefont{Liu et~al.}(2018)\citenamefont{Liu, Sun, Mao, Zhu, Ren,
  Zhou, Wang, Singh, Sui, Chu et~al.}}]{liu2018phase}
\bibinfo{author}{\bibfnamefont{Z.}~\bibnamefont{Liu}},
  \bibinfo{author}{\bibfnamefont{J.}~\bibnamefont{Sun}},
  \bibinfo{author}{\bibfnamefont{J.}~\bibnamefont{Mao}},
  \bibinfo{author}{\bibfnamefont{H.}~\bibnamefont{Zhu}},
  \bibinfo{author}{\bibfnamefont{W.}~\bibnamefont{Ren}},
  \bibinfo{author}{\bibfnamefont{J.}~\bibnamefont{Zhou}},
  \bibinfo{author}{\bibfnamefont{Z.}~\bibnamefont{Wang}},
  \bibinfo{author}{\bibfnamefont{D.~J.} \bibnamefont{Singh}},
  \bibinfo{author}{\bibfnamefont{J.}~\bibnamefont{Sui}},
  \bibinfo{author}{\bibfnamefont{C.-W.} \bibnamefont{Chu}},
  \bibnamefont{et~al.}, \bibinfo{journal}{Proceedings of the National Academy
  of Sciences} \textbf{\bibinfo{volume}{115}}, \bibinfo{pages}{5332}
  (\bibinfo{year}{2018}).

\bibitem[{\citenamefont{Wdowik et~al.}(2014)\citenamefont{Wdowik, Parlinski,
  Rols, and Chatterji}}]{wdowik2014soft}
\bibinfo{author}{\bibfnamefont{U.~D.} \bibnamefont{Wdowik}},
  \bibinfo{author}{\bibfnamefont{K.}~\bibnamefont{Parlinski}},
  \bibinfo{author}{\bibfnamefont{S.}~\bibnamefont{Rols}}, \bibnamefont{and}
  \bibinfo{author}{\bibfnamefont{T.}~\bibnamefont{Chatterji}},
  \bibinfo{journal}{Physical Review B} \textbf{\bibinfo{volume}{89}},
  \bibinfo{pages}{224306} (\bibinfo{year}{2014}).

\end{thebibliography}


\begin{references}
\bibitem{R1}
G. Tan, L.D. Zhao, M.G. Kanatzidis, Rationally Designing High-Performance Bulk Thermoelectric Materials, Chemical Reviews. 116 (2016) 12123–12149. https://doi.org/10.1021/acs.chemrev.6b00255.
\bibitem{R2}
J. Mao, H. Zhu, Z. Ding, Z. Liu, G. A. Gamage, G.Chen, Z. Ren, High Thermoelectric Cooling Performance of N-Type Mg$_3$Bi$_2$-Based Materials. Science 2019, 365 (6452), 495–498. https://doi.org/10.1126/science.aax7792.
\bibitem{R4}
Y. Xiao, H. Wu, J. Cui, D. Wang, L. Fu, Y. Zhang, Y. Chen, J. He, S.J. Pennycook, L.D. Zhao, Realizing high performance n-type PbTe by synergistically optimizing effective mass and carrier mobility and suppressing bipolar thermal conductivity, Energy Environ. Sci.,  11 (2018) 2486–2495. https://doi.org/10.1039/c8ee01151f.
\bibitem{R5}
L. Yang, Z.G. Chen, M.S. Dargusch, J. Zou, High Performance Thermoelectric Materials: Progress and Their Applications, Adv. Energy Mater. 8 (2018) 1701797. https://doi.org/10.1002/aenm.201701797.
\bibitem{R6}
T. Mori, Novel Principles and Nanostructuring Methods for Enhanced Thermoelectrics, Small. 13 (2017) 1702013. https://doi.org/10.1002/smll.201702013.
\bibitem{R7}
L. Xie, Y. Chen, R. Liu, E. Song, T. Xing, T. Deng, Q. Song, J. Liu, R. Zheng, X. Gao, S. Bai, L. Chen, Stacking faults modulation for scattering optimization in GeTe-based thermoelectric materials, Nano Energy. 68 (2020) 104347. https://doi.org/10.1016/j.nanoen.2019.104347.
\bibitem{R8}
Y. Yu, M. Cagnoni, O. Cojocaru‐Mirédin, M. Wuttig, Chalcogenide Thermoelectrics Empowered by an Unconventional Bonding Mechanism, Adv. Func. Mater. 30 (2020) 1904862. https://doi.org/10.1002/adfm.201904862.
\bibitem{R8a}
D. Sivaprahasam, S.B. Chandrasekhar, S. Kashyap, A. Kumar, R. Gopalan, Thermal conductivity of nanostructured Fe$_{0.04}$Co$_{0.96}$Sb$_3$ skutterudite, Materials Letters. 252 (2019) 231–234. https://doi.org/10.1016/j.matlet.2019.05.140.
\bibitem{R8b}
A. Kumar, A. Kosonowski, P. Wyzga, K.T. Wojciechowski, Effective thermal conductivity of SrBi$_4$Ti$_4$O$_{15}$-La$_{0.7}$Sr$_{0.3}$MnO$_3$ oxide composite: Role of particle size and interface thermal resistance, J. Eur. Cer. Soc. 41 (2021) 451–458. https://doi.org/10.1016/j.jeurceramsoc.2020.08.069.
\bibitem{R8c}
A. Kumar, K.T. Wojciechowski, Effect of interface resistance on thermoelectric properties in (1-x)La$_{0.95}$Sr$_{0.05}$Co$_{0.95}$Mn$_{0.05}$O$_3$/(x)WC composite, J. Eur. Ceram. Soc. 42 (10) (2022), 4227-4232. https://doi.org/10.1016/j.jeurceramsoc.2022.03.062
\bibitem{R8d}
A. Kosonowski, A. Kumar, T. Parashchuk, R. Cardoso-Gil, K.T. Wojciechowski, Thermal conductivity of PbTe–CoSb$_3$ bulk polycrystalline composite: role of microstructure and interface thermal resistance, Dalton Trans. 50 (2021) 1261–1273. https://doi.org/10.1039/D0DT03752D.
\bibitem{R8dd}
A. Kosonowski, A. Kumar, K. Wolski, S. Zapotoczny, K. T. Wojciechowski, Origin of electrical contact resistance and its dominating effect on electrical conductivity in PbTe/CoSb$_3$ composite, J. Eur. Ceram. Soc. 42(6) (2022) 2844-2852. https://doi.org/10.1016/j.jeurceramsoc.2022.01.049 
\bibitem{R8e}
B. Jiang, Y. Yu, J. Cui, X. Liu, L. Xie, J. Liao et al. High-entropy-stabilized chalcogenides with high thermoelectric performance. Science 371 (2021) 830–4. https://doi.org/10.1126/science.abe1292.
\bibitem{R8f}
A. Kumar, D. Dragoe, D. Berardan, and N. Dragoe, Thermoelectric Properties of High-Entropy Rare-Earth Cobaltates, Journal of Materiomics 9 (2023) 191-196. https://doi.org/10.1016/j.jmat.2022.08.001
\bibitem{R9}
Y. Pei, A. LaLonde, S. Iwanaga, G.J. Snyder, High thermoelectric figure of merit in heavy hole dominated PbTe, Energy Environ. Sci. 4 (2011) 2085. https://doi.org/10.1039/c0ee00456a.
\bibitem{R10}
L.-D. Zhao, S.-H. Lo, Y. Zhang, H. Sun, G. Tan, C. Uher, C. Wolverton, V.P. Dravid, M.G. Kanatzidis, Ultralow thermal conductivity and high thermoelectric figure of merit in SnSe crystals, Nature 508 (2014) 373–377. https://doi.org/10.1038/nature13184.
\bibitem{R10a}
Y. Tang, Z. M. Gibbs, L. A. Agapito, G. Li, H-S. Kim, M. B. Nardelli, S. Curtarolo, G. J. Snyder. Convergence of multi-valley bands as the electronic origin of high thermoelectric performance in CoSb$_3$ skutterudites. Nat. Mater. 14 (2015) 1223–1228. 
\bibitem{R10b}
B. Hinterleitner, et al. Termoelectric performance of a metastable thin-flm Heusler alloy. Nature 576 (2019) 85–90. 
\bibitem{R11}
S. Perumal, S. Roychowdhury, K. Biswas, High performance thermoelectric materials and devices based on GeTe, J. Mater. Chem. C 4 (2016) 7520–7536. https://doi.org/10.1039/C6TC02501C.
\bibitem{R12}
D.G. Cahill, R.O. Pohl, Heat flow and lattice vibrations in glasses, Sol. Stat. Comm. 70 (1989) 927–930. https://doi.org/10.1016/0038-1098(89)90630-3.
\bibitem{R13}
Y. Pei, G. Tan, D. Feng, L. Zheng, Q. Tan, X. Xie, S. Gong, Y. Chen, J. Li, J. He, M.G. Kanatzidis, L. Zhao, Integrating Band Structure Engineering with All‐Scale Hierarchical Structuring for High Thermoelectric Performance in PbTe System, Adv. Energy Mater. 7 (2017) 1601450. https://doi.org/10.1002/aenm.201601450.
\bibitem{R14}
J. Mao, Y. Wu, S. Song, Q. Zhu, J. Shuai, Z. Liu, Y. Pei, Z. Ren, Defect Engineering for Realizing High Thermoelectric Performance in n-Type Mg$_3$Sb$_2$--Based Materials, ACS Energy Lett. 2 (2017) 2245–2250. https://doi.org/10.1021/acsenergylett.7b00742.
\bibitem{R15}
A. Kumar, P. Bhumla, A. Kosonowski, K. Wolski, S. Zapotoczny, S. Bhattacharya, K. Wojciechowski, Synergistic effect of workfunction and acoustic impedance mismatch for improved thermoelectric performance in GeTe-WC composite, ACS Appl. Mater. Interfaces 14 (39) (2022) 44527-44538 https://doi.org/10.1021/acsami.2c11369
\bibitem{R17}
H. Liu, Z. Chen, J. Tang, Y. Zhong, X. Guo, F. Zhang, R. Ang, High Quality Factor Enabled by Multiscale Phonon Scattering for Enhancing Thermoelectrics in Low-Solubility n-Type PbTe–Cu$_2$Te Alloys, ACS App. Mater. Interfaces 12 (2020) 52952–52958. https://doi.org/10.1021/acsami.0c16431.
\bibitem{R18}
T. Zhu, Y. Liu, C. Fu, J.P. Heremans, J.G. Snyder, X. Zhao, Compromise and Synergy in High‐Efficiency Thermoelectric Materials, Adv. Mater. 29 (2017) 1605884. https://doi.org/10.1002/adma.201605884.
\bibitem{R19}
X. Zhang, J. Li, X. Wang, Z. Chen, J. Mao, Y. Chen, Y. Pei, Vacancy Manipulation for Thermoelectric Enhancements in GeTe Alloys, J. Am. Chem. Soc. 140 (2018) 15883–15888. https://doi.org/10.1021/jacs.8b09375.
\bibitem{R20}
X. Zhang, Z. Bu, S. Lin, Z. Chen, W. Li, Y. Pei, GeTe Thermoelectrics, Joule. 4 (2020) 986–1003. https://doi.org/10.1016/j.joule.2020.03.004.
\bibitem{R21}
D. Wu, L.-D. Zhao, S. Hao, Q. Jiang, F. Zheng, J.W. Doak, H. Wu, H. Chi, Y. Gelbstein, C. Uher, C. Wolverton, M. Kanatzidis, J. He, Origin of the High Performance in GeTe-Based Thermoelectric Materials upon Bi$_2$Te$_3$ Doping, J. Am. Chem. Soc. 136 (2014) 11412–11419. https://doi.org/10.1021/ja504896a.
\bibitem{R22}
S. Perumal, P. Bellare, U.S. Shenoy, U. v. Waghmare, K. Biswas, Low Thermal Conductivity and High Thermoelectric Performance in Sb and Bi Codoped GeTe: Complementary Effect of Band Convergence and Nanostructuring, Chem. Mater. 29 (2017) 10426–10435. https://doi.org/10.1021/acs.chemmater.7b04023.
\bibitem{R23}
L. Yue, W. Cui, S. Zheng, Y. Wu, L. Wang, P. Bai, X. Dong, Band Engineering and Thermoelectric Performance Optimization of p-Type GeTe-Based Alloys through Ti/Sb Co-Doping, J. Phys. Chem. C 124 (2020) 5583–5590. https://doi.org/10.1021/acs.jpcc.0c00045.
\bibitem{R24}
T. Mori, D. Berthebaud, T. Nishimura, A. Nomura, T. Shishido, K. Nakajima, Effect of Zn doping on improving crystal quality and thermoelectric properties of borosilicides, Dalton Trans. 39 (2010) 1027–1030. https://doi.org/10.1039/B916028K.
\bibitem{R25}
J.K. Lee, M.W. Oh, B.S. Kim, B.K. Min, H.W. Lee, S.D. Park, Influence of Mn on crystal structure and thermoelectric properties of GeTe compounds, Elec. Mater. Lett. 10 (2014) 813–817. https://doi.org/10.1007/s13391-014-4149-8.
\bibitem{R26}
A. Kumar, P. Bhumla, T. Parashchuk, S. Baran, S. Bhattacharya, K.T. Wojciechowski, Engineering Electronic Structure and Lattice Dynamics to Achieve Enhanced Thermoelectric Performance of Mn–Sb Co-Doped GeTe, Chem. Mater. 33 (2021) 3611–3620. https://doi.org/10.1021/acs.chemmater.1c00331.
\bibitem{R28}
J. Dong, F.-H. Sun, H. Tang, J. Pei, H.-L. Zhuang, H.-H. Hu, B.-P. Zhang, Y. Pan, J.-F. Li, Medium-temperature thermoelectric GeTe: vacancy suppression and band structure engineering leading to high performance, Energy Environ. Sci. 12 (2019) 1396–1403. https://doi.org/10.1039/C9EE00317G.
\bibitem{R27}
J. Shuai, Y. Sun, X. Tan, T. Mori, Manipulating the Ge Vacancies and Ge Precipitates through Cr Doping for Realizing the High‐Performance GeTe Thermoelectric Material, Small. 16 (2020) 1906921. https://doi.org/10.1002/smll.201906921.
\bibitem{hohenberg1964inhomogeneous}
P. Hohenberg; W. Kohn, Inhomogeneous Electron Gas. Phys. Rev. 1964, 136, No. B864.
\bibitem{kohn1965self}
Kohn, W.; Sham, L. J. Self-Consistent Equations Including Exchange and Correlation Effects. Phys. Rev. 1965, 140, No. A1133
\bibitem{kresse1996efficiency}
G. Kresse, J. Furthmuller, Efficiency of Ab-initio total energy calculations for metals and semiconductors using a plane-wave basis set. Comput. Mater. Sci. 1996, 6, 15-50. 
\bibitem{kresse1999ultrasoft}
Kresse, G.; Joubert, D. From Ultrasoft Pseudopotentials to the Projector Augmented-Wave Method. Phys. Rev. B 1999, 59, No. 1758.
\bibitem{perdew1996generalized}
J. P. Perdew, K. Burke, M. Ernzerhof, Generalized gradient approximation made simple, Phys. Rev. Lett. 77(18) (1996) 3865.
\bibitem{momma2011vesta}
K. Momma, F. Izumi, VESTA 3 for three-dimensional visualization of crystal, volumetric and morphology data, J. Appl. Cryst. 44(6) (2011) 1272--1276.
\bibitem{parlinski1997first}
K. Parlinski, Z. Li, Y. Kawazoe. First-principles determination of the soft mode in cubic ZrO$_2$, Phys. Rev. Lett. 78(21) (1997) 4063.
\bibitem{togo2015first}
A. Togo, I. Tanaka, First principles phonon calculations in materials science, Scr. Mater. 108 (2015) 1--5. 
 \bibitem{togo2008first}
  	A. Togo, F. Oba, I. Tanaka, First-principles calculations of the ferroelastic transition between rutile-type and CaCl$_2$-type SiO$_2$ at high pressures, Phys. Rev. B 78(13) (2008) 134106. 
   \bibitem{madsen2006boltztrap}
G.  Madsen, D. J. Singh, BoltzTraP. A code for calculating band-structure dependent quantities, Computer Physics Communications, 175(1) (2006) 67--71.
\bibitem{shannon}
RD. Shannon, Revised effective ionic radii and systematic studies of interatomic distances in halides and chalcogenides. Acta Cryst. A 32 (1976) 751–67. https://doi.org/10.1107/S0567739476001551.
\bibitem{structural}
S. Perumal, M. Samanta, T. Ghosh, U.S. Shenoy, A. Bohra, S. Bhattacharya, A. Singh, U V Waghmare, K. Biswas, Realization of High Thermoelectric Figure of Merit in GeTe by Complementary Co-Doping of Bi and In. Joule. 3 (2019) 2565--2580.
%\bibitem{Li}
%J. Li, C. Zhang, Y. Feng, C. Zhang, Y. Li, L. Hu, W. Ao, F. Liu, Effects on phase transition and thermoelectric properties in the Pb-doped GeTe-Bi$_2$Te$_3$ alloys with thermal annealing, Journal of Alloys and Compounds. 808 (2019) 151747. https://doi.org/10.1016/j.jallcom.2019.151747.
\bibitem{R31}
D. Wu, L. Xie, X. Xu, J. He, High Thermoelectric Performance Achieved in GeTe–Bi$_2$Te$_3$ Pseudo‐Binary via Van der Waals Gap‐Induced Hierarchical Ferroelectric Domain Structure, Adv. Func. Mater. 29 (2019) 1806613. https://doi.org/10.1002/adfm.201806613.
\bibitem{R32}
J. Li, X. Zhang, Z. Chen, S. Lin, W. Li, J. Shen, I.T. Witting, A. Faghaninia, Y. Chen, A. Jain, L. Chen, G.J. Snyder, Y. Pei, Low-Symmetry Rhombohedral GeTe Thermoelectrics, Joule. 2 (2018) 976–987. https://doi.org/10.1016/j.joule.2018.02.016.
\bibitem{R33}
Yu.I. Ravich, B.A. Efimova, I.A. Smirnov, Semiconducting Lead Chalcogenides, 1970. https://doi.org/10.1007/978-1-4684-8607-0.
\bibitem{R34}
N.K. Abrikosov, V.F. Bankina, L. v. Poretskaya, L.E. Shelimova, E. v. Skudnova, Semiconducting II–VI, IV–VI, and V–VI Compounds, 1969. https://doi.org/10.1007/978-1-4899-6373-4.
\bibitem{mobility}
G. J. Snyder, A. H. Snyder, M. Wood, R. Gurunathan, B. H. Snyder, C. Niu, Weighted Mobility, Adv. Mater. 32 (2020) 2001537.  https://doi.org/10.1002/adma.202001537
\bibitem{liu2018phase}
Z. Liu, J. Sun, J. Mao, H. Zhu, W. Ren, J. Zhou, Z, Wang, D. J. Singh, J. Sui, C--W, Chu, Z. Ren, Phase-transition temperature suppression to achieve cubic GeTe and high thermoelectric performance by Bi and Mn codoping, PNAS, 115 (21), (2018) (5332--5337).
\bibitem{phonon}
S. Perumal, S. Roychowdhury, D.S. Negi, R. Datta, K. Biswas, High Thermoelectric Performance and Enhanced Mechanical Stability of p-type Ge$_{1-x}$Sb$_x$Te, Chem. Mater. 27 (2015) 7171–7178. https://doi.org/10.1021/acs.chemmater.5b03434.
\bibitem{wdowik2014soft}
U. D. dowik, K. Parlinski, S. Rols, T. Chatterji, Soft-phonon mediated structural phase transition in GeTe, Phys. Rev. B, 89(22) (2014) 224306.
\bibitem{zT1}
S. Roychowdhury, M. Samanta, S. Perumal, K. Biswas, Germanium Chalcogenide Thermoelectrics: Electronic Structure Modulation and Low Lattice Thermal Conductivity, Chem. Mater. 30 (2018) 5799–5813. https://doi.org/10.1021/acs.chemmater.8b02676.
\bibitem{zT3}
S. Perumal, S. Roychowdhury, K. Biswas, Reduction of thermal conductivity through nanostructuring enhances the thermoelectric figure of merit in Ge$_{1-x}$Bi$_x$Te, Inorg. Chem. Front. 3 (2016) 125–132. https://doi.org/10.1039/c5qi00230c.
\bibitem{zT4}
Z. Zheng et al., Rhombohedral to Cubic Conversion of GeTe via MnTe Alloying Leads to Ultralow Thermal Conductivity, Electronic Band Convergence, and High Thermoelectric Performance, J. Am. Chem. Soc. 140 (2018) 2673–2686. https://doi.org/10.1021/jacs.7b13611.
\bibitem{zT6}
Z.W. Lu, J.Q. Li, C.Y. Wang, Y. Li, F.S. Liu, W.Q. Ao, Effects of Mn substitution on the phases and thermoelectric properties of Ge$_{0.8}$Pb$_{0.2}$Te alloy, J. Alloys Compd. 621 (2015) 345–350. https://doi.org/10.1016/j.jallcom.2014.09.198.
\bibitem{zT7}
L. Wu, X. Li, S. Wang, T. Zhang, J. Yang, W. Zhang, L. Chen, J. Yang, Resonant level-induced high thermoelectric response in indium-doped GeTe, NPG Asia Mater. 9 (2017) 1–7. https://doi.org/10.1038/am.2016.203.
\end{references}
\end{document}